\documentclass[aps,prb,twocolumn,footinbib,superscriptaddress]{revtex4-2}
\usepackage{feynmp}

\usepackage{amsmath}
\usepackage{amssymb}
\usepackage{bbm}
\usepackage{braket}
\usepackage{xcolor}
\usepackage{pifont}
\usepackage{slashed}
\usepackage[mathscr]{euscript}
\usepackage[shortlabels]{enumitem}
\usepackage{bm}
\usepackage{dsfont}
\usepackage{subfigure}
\allowdisplaybreaks

\usepackage{graphicx}
\usepackage[colorlinks=true]{hyperref}
\usepackage{comment}

\newcommand{\bsub}{\begin{subequations}}
\newcommand{\esub}{\end{subequations}}

\definecolor{wrongultramarine}{rgb}{1,0.5,0}

\newcommand{\be}{\begin{equation}}
\newcommand{\ee}{\end{equation}}
\newcommand{\beq}{\begin{eqnarray}}
\newcommand{\eeq}{\end{eqnarray}}
\newcommand{\ba}{\[\begin{aligned}}
\newcommand{\ea}{\end{aligned}\]}
\newcommand{\bal}{\begin{aligned}}
\newcommand{\eal}{\end{aligned}}

\newcommand{\tr}{{\rm tr\,}}

\renewcommand{\epsilon}{\varepsilon}

\newcommand{\ii}{\mathrm{i}}

\hypersetup{
    bookmarks=true,         
    unicode=false,          
    pdftoolbar=true,        
    pdfmenubar=true,        
    pdffitwindow=false,     
    pdfstartview={FitH},    
    pdfsubject={},   
    pdfcreator={},   
    pdfproducer={}, 
    pdfkeywords={} {} {}, 
    pdfnewwindow=true,      
    colorlinks=true,       
    linkcolor=blue, 
    citecolor=blue,        
    filecolor=magenta,      
    urlcolor=blue           
}

\newcommand{\cl}{\mathsf{cl}}
\newcommand{\q}{\mathsf{q}}

\newcommand{\Nabla}{\bm{\nabla}}
\newcommand{\T}{\mathsf{T}}
\newcommand{\sigh}{\hat{\sigma}}
\newcommand{\tauh}{\hat{\tau}}
\newcommand{\intl}[1]{\int\limits_{#1}}
\newcommand{\parr}{\partial}

\DeclareMathOperator{\trr}{tr}
\newcommand{\ord}[1]{\bm{\mathit{O}}\left(#1\right)}

\usepackage{color}

\newcommand{\bz}{\mathsf{BZ}}
\newcommand{\te}{\tilde{\varepsilon}}
\newcommand{\Mp}{\hat{M}_P}

\definecolor{DarkBlue}{rgb}{.3, 0, 0.85}

\definecolor{OliveGreen}{cmyk}{0.64, 0, 0.95, 0.40}

\begin{document}

\title{Free-Fermion  Measurement-Induced Volume- to Area-Law Entanglement Transition in the Presence of Fermion Interactions}

\begin{abstract}
At a generic volume- to area-law entanglement transition in a many-body system, quantum chaos is arrested. We argue that this tends to imply the vanishing of a certain ``mass'' term in the field theory of the measurement-induced phase transition (MIPT) for monitored, interacting fermions. To explore this idea, we consider the MIPT with no conserved quantities that describes 1D  monitored, interacting Majorana fermions in class DIII. 
This is the \emph{most general} problem of interacting fermions with weak fermion parity measurements.
Without interactions, it is known that a noninteracting MIPT separates the area-law phase from a log-enhanced ``thermal metal'' phase at sufficiently weak monitoring. We conjecture that the 
MIPT \emph{with interactions} is the same as the
\emph{noninteracting} 
one in this case; the volume-law phase arises through the dangerously irrelevant mass. 
The physical picture is that the mass represents a local Fermi's golden rule interparticle scattering rate density that is tantamount to the entangling rate density. The latter must vanish continuously at a \emph{continuous} MIPT. 
On the other hand, the field theory capturing the MIPT for monitored fermions with additional continuous symmetries is expected to be different, because the interactions introduce additional terms associated to conserved Noether currents.  
We propose numerical tests of our conjecture. 
In addition, we analytically identify a candidate noninteracting critical point representing the MIPT, using a controlled $\epsilon$-expansion.
\end{abstract}

\author{Matthew S. Foster}
\email{matthew.foster@rice.edu}
\affiliation{Department of Physics and Astronomy, Rice University, Houston, Texas 77005, USA}
\author{Haoyu Guo}
\affiliation{Department of Physics, Cornell University, Ithaca, New York 14853, USA}
\author{Chao-Ming Jian }
\affiliation{Department of Physics, Cornell University, Ithaca, New York 14853, USA}
\author{Andreas W. W. Ludwig}
\affiliation{Department of Physics, University of California, Santa Barbara, California 93106, USA}

\date{\today}

\maketitle


\section{Introduction \label{sec:Intro}}

Measurement-induced phase transitions (MIPTs) \cite{li2018quantum,li2019measurement,skinner2019measurement,chan2019unitary,fisher2023random,PotterVasseurReview2021}
represent a fascinating class of far-from-equilibrium quantum localization phenomena. The entanglement structure of a quantum many-body state subject to both generic (``quantum-chaotic'') unitary evolution and measurements is determined by the competition of these, and has drawn intense interest from physicists working in quantum information, quantum dynamics, and many-body theory \cite{zhou2019emergent,cao2019entanglement,VasseurRTN2019,zabalo2020critical,choi_2020,Jian20,Bao20,nahum2020entanglement,lang2020entanglement,gullans2020dynamical,gullans2020scalable,fidkowskiHaahHastingsHowDynamicalMemoriesForget-arXiv2008.10611,lavasani2021measurement,Sang2021LoopModel,Turkeshi2020MIPT2D,Zabalo2021,LiChenLudwigFisher2021,NahumRoySkinnerRuhmanAlltoAll2021,AlbertonBuchholdDiehlFermionPRL2021,MBuchhold2021,CMJian2022,Agrawal22,Barratt_U1_FT,LearnabilityPhysRevLett.129.200602,li2023cross,MajidyAgrawalGopalakrishnanPotterVasseurHalpern2023,LiVijayPolymer2023entanglement,Jian23,MFava2023,Poboiko23,Chahine23,NahumWiese,PiroliLiVasseurNahumControl2023,LiVasseurFisherLudwig,KumarKemalChakrabortyLudwigGopalakrishnanPixleyVasseur,VidalPotterVasseur2024,LovasAgrawalVijayBoundaryDiss2024,Mirlin2024_Above1D,Fava24,MerrittFidkowski2023,Lumia2024,YoheiAshida2020}.

Recent analytical progress includes the development of effective field theories for the monitored dynamics and MIPT of interacting fermions \cite{Guo24,Poboiko24,Tiutiakina24}
\footnote{See also 
Refs.~\cite{NahumRoySkinnerRuhmanAlltoAll2021,NahumWiese}
for a different approach.}. 
These theories take a form similar to the \emph{replica} nonlinear sigma models familiar from the physics of 
single-particle Anderson localization 
\cite{Zirnbauer96,Evers08}, and were
first pioneered for the monitored dynamics of noninteracting fermions \cite{Jian23,MFava2023,Poboiko23,Fava24}. 
However, (i) different
from 
localization, the relevant physics arises in the 
replica number
$R \rightarrow 1$ limit \cite{Jian20,Bao20}, 
and  (ii) the volume-law entanglement phase is enabled by an {\it additional}
replica-anisotropic
``mass'' term induced by the interactions \cite{Guo24,Poboiko24,Tiutiakina24}: 
The sigma models describing monitored, \emph{noninteracting} fermions possess a 
large continuous replica rotation symmetry that is explicitly broken down by the mass term to a subgroup
of discrete replica permutations.

In this work, we reconsider the generic monitored dynamics of
interacting fermions in one spatial dimension, without any conserved quantities.
These systems can be treated as the monitored circuits in symmetry class DIII~\cite{CMJian2022,Jian23,MFava2023}.
(``Class DIII''
refers to the noninteracting limit of these circuits. A more rigorous formulation of the interacting symmetry class is given in Ref.~\cite{Jian25}.) 

Using the non-linear sigma-model description augmented by interactions,
we conjecture that the critical point of the MIPT in the weakly interacting class-DIII monitored circuit is identical to that of the \emph{noninteracting} transition \cite{Jian23,MFava2023}. 
This is because the aforementioned mass term becomes
\emph{dangerously irrelevant} at the critical point of the interacting circuit. The mass
term enables the appearance of volume-law entanglement scaling when the interacting circuit deviates from the critical point.  
The idea is that the symmetry-breaking mass term, originating from the presence of interactions, encodes dynamical quantum chaos facilitated by interparticle scattering, as manifested in the volume law phase. This chaos is suppressed to subextensive levels in the area-law phase, in which interactions are immaterial.  
Note that our arguments will not apply to cases with extra 
conservation laws, such as conserved U(1) charges.
Continuous symmetries
preserved in every quantum trajectory can enable further interaction terms in the effective field theory. Such terms (which also break the continuous free-fermion symmetry) were identified for class AIII in Ref.~\cite{Guo24}, where interactions are required for the \emph{existence}
of a MIPT
\cite{Poboiko23,Fava24}.

While monitored fermionic dynamics without conserved charges are interesting in their own right, they are also closely related to the decoding problem of surface codes with coherent errors \cite{BravyiCoherentError2018,Venn2023,BehrendPRR2024,YangLudwigJian2026,YanBaoVijay2026}. In fact, the decodability transitions of the latter system under coherent errors are dual to the MIPTs in the monitored circuits on a Majorana fermion chain. The form of the coherent error controls whether the interaction is present in the dual monitored circuits. 
For the square-lattice surface code, it turns out these dual monitored circuits belong to symmetry class D in the noninteracting limit, which can be viewed as a class-DIII monitored circuit with an extra symmetry or constraints \cite{CMJian2022,Jian23,MFava2023,wang2025decoherenceinducedselfdualcriticalitytopological,GRL2001,Jian25}.
The numerical simulations of these dual class-D monitored circuits with and without interactions \cite{Venn2023,BehrendPRR2024} show a tantalizing sign of impeded development of volume-law entanglement scaling even if interactions are present, which might result from strong renormalization of the interaction-induced replica-anisotropic mass term. A similar renormalization effect of the mass term is indeed the focus of this current work. On the other hand, for the decoding problems in the honeycomb-lattice and triangular-lattice surface code, the detailed forms of coherent errors determine whether the dual circuit belongs to class DIII or D and whether interactions are present \cite{YanBaoVijay2026,YangLudwigJian2026}. For this work, we focus on   
interacting symmetry-class DIII
fermionic monitored circuits
which are fermionic monitored circuits without extra constraints other than global fermion-parity conservation.
Our study here will set the stage for investigating interacting class-D circuits 
and surface code decoding problems dual to Majorana monitored circuits 
in the future. 

In this article, 
we 
consider
the field theory of the interaction-deformed class-DIII MIPT within the Keldysh formalism by extending the analysis in Refs.~\cite{Guo24,Poboiko24} (which addressed symmetry class AIII). 
We provide a general discussion of symmetries, and basic properties of renormalization group flows.
This discussion motivates our conjecture on the nature of the MIPT in weakly interacting monitored circuits.
We then propose 
numerical tests for our conjecture on the nature of the interaction-deformed class-DIII MIPT and the effect of the aforementioned dangerously irrelevant term at the critical point.

In addition, expanding on Refs.~\cite{Jian23,MFava2023}, we develop an 
analytical approach to the theory of the 
class-DIII MIPT by exploiting a controlled $
R = 2 - \epsilon < 2$ expansion. This expansion is motivated by pioneering work
due to 
Fu and Kane \cite{FuKane12} 
in the context of 2D Anderson localization with spin-orbit coupling, and later adapted~\cite{Jian23,MFava2023}
to the problem of noninteracting monitored circuits in class DIII.
This
approach, although not fully controlled, 
predicts the correct topology of the noninteracting
RG flows. 
We introduce a \emph{controlled} $\epsilon$-expansion 
to access the interaction-deformed class-DIII MIPT. 
We find a {\it noninteracting} MIPT where the symmetry-breaking mass term is a dangerously irrelevant perturbation to this fixed point: The volume-law phase can be reached from the MIPT by RG flow in the presence of arbitrarily weak interactions. 
The RG flow is summarized in Fig.~\ref{Fig--Flow}.

\subsection{Outline}

The outline of this paper is as follows. The setup and physical picture is articulated in Sec.~\ref{sec:results}, followed
by proposed numerical tests and an overview of the analytical results for the MIPT. 
Technical details are summarized in Sec.~\ref{sec:details}. We conclude and enumerate key open questions in Sec.~\ref{sec:conclusion}. Additional technical components of our analysis are relegated to the Appendices.


\begin{figure}[b!]
\centering
\includegraphics[width=0.45\textwidth]{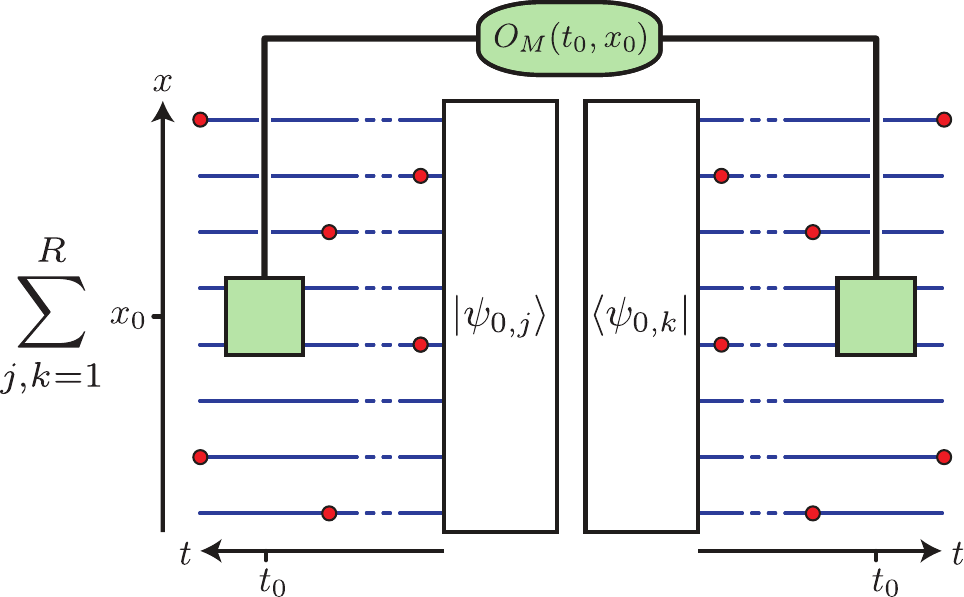}
\caption{The mass operator for the NLsM encodes interparticle scattering,
and acts like an \emph{entanglement-sector jump term.}
Green boxes represent the effects of 2-particle interaction operators applied simultaneously to the ``ket'' and ``bra'' sides of the density matrix; red dots indicate measurements. 
}
\label{fig:masses}
\end{figure}

\section{Non-interacting MIPT for monitored, interacting fermions \label{sec:results}}

\subsection{Field theory and nature of the MIPT}

For concreteness, we describe the physics in terms of the monitored dynamics of a particular 1+1-D mean-field superconductor model, although the field theory defined below captures universal aspects of any sufficiently local system in class DIII. 
We employ the superconductor language mainly because it allows a recognizable action and symmetry analysis
(see Appendix~\ref{app:sym})
that is familiar from the different problem of interacting, disordered fermions in equilibrium \cite{Liao17}. We emphasize, however, that monitored dynamics induce a scrambling of fermions across the band (``heating effect'' \cite{Poboiko23,Guo24,Poboiko24}), and are unrelated to the low-temperature physics of ordinary equilibrium superconductivity. Our results are most relevant to engineered quantum platforms such as quantum circuit dynamics without U(1) conservation and quantum error correction in the surface code, as described in Sec.~\ref{sec:Intro}.

The kinetic term of the superconductor model consists of spinless fermions hopping on a 1D lattice with normal and pairing terms, see Eq.~(\ref{H0}). We monitor the local fermion density operator $c_x^\dagger c_x$. In addition, local density-density interactions are incorporated with a coupling strength $U$. 
Provided the interactions are sufficiently short-ranged, the precise form is not important for the coarse-grained effective field theory.

Following 
Refs.~\cite{Poboiko23,Fava24,Guo24,Poboiko24,Jian23,MFava2023},
we formulate a Keldysh path integral for the system. We replicate the field theory $R$ times in order to  encode observables nonlinear in the density matrix and to enable averaging over quantum trajectories. 
One must take the limit of $R \rightarrow 1$ replicas at the end of the calculation 
to enforce the Born rule
\cite{Jian20,Bao20}. We perform the trace-log expansion around the measurement-induced saddle-point and arrive at the 
entanglement field-theory action \cite{Guo24,Poboiko24} 
(see also Appendix~\ref{app:sym})
\begin{align}\label{S}
    \!
    S 
    = 
    \int d t \, d x
    \,
    \left\{
        \frac{\lambda}{16}
        \trr\left[\Nabla \hat{X}^T \cdot \Nabla \hat{X} \right]
        -
        \frac{M}{16}
        \,
        \mathcal{O}_M(t,x)
    \right\}\!.\!
\end{align}
The field variable $\hat{X} \rightarrow X_{j,k}$ is an SO($R$) group element matrix~\cite{Jian23,MFava2023}
with
replica indices $j,k \in \{1,2,\ldots,R\}$, 
which satisfies the constraint
$\sum_{j=1}^R X_{j,i} \, X_{j,k} = \delta_{i,k}$. 
In Eq.~(\ref{S}),
$\Nabla \equiv \{\parr_t,\parr_{x}\}$ is the \emph{Euclidean} spacetime gradient operator,
$\lambda$ is the ``stiffness'' (inversely proportional to the measurement rate),
and $M$ is the interaction-induced ``mass'' (discussed more below).

The action in Eq.~(\ref{S}) can be obtained from
Ref.~\cite{Guo24,Poboiko24}, describing
U(1) charge-conserving monitored fermions (class AIII in the noninteracting limit),
by 
suppressing the charge degrees of freedom.
One can associate the matrix field appearing in Eq.~\eqref{S}
to fermion bilinears via
$
    X_{j,k} 
    \Leftrightarrow 
    c_{+,j} \, \bar{c}_{-,k}
$
and
$
    X^\dagger_{j,k}
    \Leftrightarrow
    c_{-,j} \, \bar{c}_{+,k},
$
where $c_{+,j}$ ($\bar{c}_{-,k}$) denotes an annihilation (creation) field on the forward (backward) Keldysh contour \cite{Guo24}.

The
noninteracting theory 
in Eq.~\eqref{S} but
with $M = 0$ has SO($R$)$\times$SO($R$) symmetry, corresponding to independent replica rotations on the forward and backward Keldysh contours. 
(Compare to Ref.~\cite{Guo24}, and 
Refs.~\cite{Jian23,MFava2023}.)
The interaction operator 
$
    O_M \equiv \sum_{j,k = 1}^R X_{j k}^4 
$
breaks the symmetry down to $S_R$ $\times$ $S_R$, where $S_R \subset {\rm SO}(R)$ is the discrete permutation group on $R$ objects. The mass coupling
constant $M \propto (\lambda U)^2$ encodes the inelastic scattering rate of the fermions, and arises from an electron self-energy in the derivation of Eq.~(\ref{S}) \cite{Guo24}. 
$\mathcal{O}_M$
can be viewed as an \emph{entanglement-sector jump term},
due to the correspondence (see also Fig.~\ref{fig:masses})
\begin{align}\label{OMJump}
    \mathcal{O}_M 
    \Leftrightarrow 
    \sum_{j,k = 1}^R 
    \bar{c}_{+,j} \, c_{+,j} \, \bar{c}_{+,j} \, c_{+,j} 
    \, 
    \bar{c}_{-,k} \, c_{-,k} \, \bar{c}_{-,k} \, c_{-,k}.
\end{align}
This operator implements ``hopping'' between different many-fermion states in Fock space, and plays a similar role as inelastic processes in theories of many-body localization \cite{Altshuler97,BAA}.
Jump terms are familiar from the Markovian dynamics of the \emph{average} density matrix in open quantum systems subjected to decoherence that are described by the Lindblad equation, where such terms encode particle scattering. This is similar to the collision integral in kinetic theory; such Lindbladian dynamics can be also cast in the Keldysh formalism \cite{Sieberer16}.
By contrast, $\mathcal{O}_M$ implements Fock-space mixing in the 
replicated entanglement sector of a monitored, but otherwise isolated 
fermion system where measurement outcomes are collected but not traced out.

The dynamical phase diagram for the theory in Eq.~(\ref{S}) can be understood from a combined renormalization group (RG) and symmetry perspective.
The sigma model is weakly coupled for large $\lambda$ (rare measurements) and small $M$ (weak interactions).
In this regime the one-loop RG equations for a 2D spacetime circuit are 
\begin{align}\label{RGpert}
    \frac{d \lambda}{d \ln L}
        =
        \frac{2 - R}{\pi},
\quad
    \frac{d M}{d \ln L}
    =
    (2 - \Delta_M) M,
\end{align}
where $L$ is the linear system size and
where 
$
    \Delta_M
    =
    4(R+2)/\pi \lambda
$
is the mass-operator scaling dimension.
In the $R \rightarrow 1$ limit, 
Eq.~(\ref{RGpert}) shows that the stiffness 
$\lambda$ (whose bare value is inversely proportional to the measurement rate)
tends to \emph{increase}, while the mass
$M$, a measure of 
interaction-mediated scattering
relative to the measurement rate,
rapidly grows to large values. 

For non-vanishing $M$, Eq.~(\ref{S}) has 
$S_R$$\times$$S_R$ symmetry under $\hat{X} \rightarrow \hat{W}_+ \hat{X} \hat{W}_-$,
where $\hat{W}_{\pm}$ represent independent permutations acting on forward- and backward-branch fermions.
The RG flow $M \rightarrow \infty$ of the mass
induces spontaneous symmetry breaking, so that
a non-vanishing expectation value $\langle X_{ij} \rangle \neq 0$ appears,
reducing the
(``strong'') symmetry
$S_R \times S_R$ of the ``mass term''
to a diagonal 
(``weak'') 
permutation subgroup 
$S_R$. This is the volume-law phase.
By contrast, sufficiently rapid measurements (small $\lambda$) prevent the spread of entanglement and should stabilize
the area-law phase (which restores $S_R$$\times$$S_R$, and is analogous to a thermally disordered paramagnet). 
In between these one expects a MIPT described by a non-unitary conformal field theory (CFT).

Many-body chaos is mediated in the volume-law phase by the mass operator;
the coupling strength $M$ determines
in particular the domain-wall tension 
that is the hallmark of
the volume-law phase \cite{Guo24,Poboiko24}. 
Dynamical chaos is arrested in the area-law phase, which 
is sub-extensively entangled. 
We therefore expect that at the MIPT the 
scaling dimension of the mass is larger than the spacetime dimension, 
$\Delta_M > 2$,
so that the mass becomes a \emph{dangerously irrelevant} perturbation to the \emph{noninteracting MIPT} critical point.
Irrelevant, because $M$ would be expected to flow
to zero in this case, but dangerously so, 
in that tuning $\lambda$ slightly beyond the critical value (into the volume-law phase)
ultimately generates a large mass; see also Fig.~\ref{Fig--Flow} for the RG flow topology.
The key idea is that the mass $M$ describes the \emph{rate density} of entanglement generation 
and this should vanish at the \emph{continuous} MIPT, given that it vanishes in the area-law phase. 
The logic is similar to that determining the validity of Fermi's golden rule: 
Upon transitioning into a phase with an effectively discrete spectrum of accessible final states, the notion of a \emph{finite scattering rate} breaks down for sufficiently weak interactions \cite{Altshuler97,BAA,Micklitz22}. A continuous transition implies a continuously vanishing rate \cite{Liao18}.

The conclusion is that the critical theory 
capturing
the
MIPT for monitored, interacting Majorana fermions in class DIII
should be the same as the noninteracting one.
The field theory describing the universality class of the noninteracting DIII monitored dynamics
was derived in Refs.~\cite{Jian23,MFava2023}.
A noninteracting MIPT was shown to separate the area-law phase from a ``thermal metal,'' with entanglement scaling logarithmically enhanced beyond the area-law in the latter \cite{Jian23,MFava2023}.
The stability of the 
thermal metal 
follows from
the weak-coupling RG flow in Eq.~(\ref{RGpert}) with $M = 0$, 
which gives $\lambda(L) \sim (1/\pi) \ln(L/a)$, with $a$ the short-distance cutoff.  
From the symmetry perspective, the mass term $M$ is 
the leading anisotropy in reducing the replica symmetry from the continuous SO$(R)$$\times$SO$(R)$ group to $S_R \times S_R$. 
If the mass term is irrelevant, when placed on the critical manifold, the interactions flow to zero, recovering an emergent SO$(R)$$\times$SO$(R)$ corresponding to the noninteracting MIPT.
In Ref.~\cite{Jian23} the noninteracting 
MIPT
was studied numerically, and universal characteristics of the (strongly coupled) noninteracting CFT were computed.

We stress that our arguments for the \emph{noninteracting} character of
the critical theory at the volume-to-area-law MIPT in the \emph{presence of interactions}
are specific to class DIII (and possibly class D).
Interacting monitored fermions with additional symmetries such as U(1) charge conservation (class AIII) admit additional \emph{interactions between Noether currents} of residual
continuous symmetries \cite{Guo24}. 
Such terms can still break the continuous replica symmetry of the non-interacting limit in the IR of the MIPT.
These are known in this case to lead to a critical theory 
that occurs only in the presence of both measurements
\emph{and non-vanishing}
interactions \footnote{In this symmetry class (AIII) there is no MIPT in the absence of interactions \cite{Poboiko23}.}.


\subsection{Proposed numerical tests \label{sec:num}}

(i): A
powerful and useful tool for determining the universality class of a given MIPT,
interacting or not, 
is the so-called ``effective central 
charge''~\cite{Zabalo2021}
$c_{\rm eff}$, 
also sometimes referred to
as the ``Casimir central charge''~\cite{wang2025decoherenceinducedselfdualcriticalitytopological,pütz2025flownishimoriuniversalityweakly}. 
It can be extracted directly from the finite-size scaling behavior of the Shannon Entropy for the measurement record 
of
the circuit on long 
cylinders~\footnote{This quantity is a generalization of the analogous quantity, introduced in a different context in 
\cite{LUDWIG1987687}, to the setting of quantum measurements.}, and has been successfully
used~\cite{Zabalo2021}
to distinguish different universality classes of MIPTs, including fully interacting cases with a qubit onsite Hilbert space 
that
possess a transition into a volume-law phase.
We propose comparing the numerically computed $c_{\rm eff}$ 
for both the noninteracting DIII MIPT \cite{Jian23,MFava2023} 
and the interacting DIII MIPT.
It may also be instructive to compare these to 
$c_{\rm eff}$ obtained in Ref.~\cite{Zabalo2021}
for the generic 
MIPT
of qubits without any symmetries.

(ii): Another useful characterization of an MIPT universality class is the universal coefficient of the logarithm of subsystem size in the von Neumann (or for that matter any $n^{\rm th}>1$ R\'enyi) entanglement entropy of the quantum state at the final circuit time. This quantity is also sometimes referred to as 
``entanglement central charge''~\cite{wang2025decoherenceinducedselfdualcriticalitytopological,pütz2025flownishimoriuniversalityweakly} $c_{\rm ent}$. It has been computed numerically in~\cite{Jian23} for the noninteracting class DIII circuit 
with the result $c_{\rm ent}=0.39\pm 0.02$ for the von Neumann entropy. 
We propose comparing $c_{\rm ent}$ at the noninteracting and interacting DIII MIPTs.

(iii): A further useful universal quantity characteristic of the universality class of an MIPT involving fermions (interacting or noninteracting) is the {\it typical} critical exponent of $G$, defined to be the square of the fermion-fermion correlation function
at the final time-slice of the circuit. Since this quantity involves the quantum-trajectory average of the logarithm of the square of the fermion correlation function, a self-averaging quantity, it is less prone to strong statistical fluctuations.
This exponent has been computed numerically for the noninteracting circuit in~\cite{Jian23}. 
We propose calculating the same exponent in the interacting case and compare these two.

(iv): A diagnostic related to, but different from item (iii) is the algebraic
decay exponent of the {\it first} moment average over quantum trajectories at the MIPT of 
$G$ 
defined in (iii), above.
In the noninteracting case, continuous replica symmetry constrains the scaling dimension of the associated fermion bilinear to unity,
see 
Sec.~\ref{sec:numdetails}
for elaboration on this point.
For the interacting class DIII, which has no 
continuous symmetries, 
an interacting MIPT would generically give a non-unity dimension for the average of $G$.
A value close to unity was 
found in the noninteracting case \cite{Jian23}.

(v): Finally, we list a numerical test not of the universality class of the MIPT, but of the presence of the volume law {\it phase}. In the volume-law phase of the 
interacting DIII circuit,
we expect to see generic quantum chaos, whose presence could be diagnosed by investigating the spectrum of the reduced density matrix of the quantum state at the final time-slice of the circuit on a finite spatial interval. One signature of many-body quantum chaos 
would be level repulsion in the spectrum of this reduced density matrix. 
Numerical feasibility for observing level repulsion in the volume-law phase of generic monitored qubit circuits was demonstrated in Ref.~\cite{ChamonMuccioloRuckensteinLevelRepulsionCircuits}.


\subsection{Analytical approach: Epsilon expansion \label{sec:analytical}}

The MIPT in the theory described by Eq.~(\ref{S}) resides in a strong-coupling regime with $\lambda \sim \ord{1}$, and this presents a calculational hurdle.
A less obvious but potentially more fundamental barrier is the fact that the target manifold in this case is the orthogonal group 
SO($R$). 
In a pioneering work \cite{FuKane12}, Fu and Kane noted that Kosterlitz-Thouless physics can come into play when $R$ is deformed through $R = 2$. They were concerned with the different problem of the 2D Anderson metal-insulator transition in the symplectic (spin-orbit) class and its sigma-model description. In order to deform $R$ to its
value of interest
(zero for localization, one for the MIPT \cite{Jian23}), we have to pass through the special case of $R = 2$, i.e.\ the classical $XY$ model. Fu and Kane argued that in order to understand the symplectic Anderson transition, one must take into account the vortex physics of the Kosterlitz-Thouless transition 
native to $R = 2$ \footnote{For both classes AII and DIII, the homotopy result 
$\Pi_1[G/H] = \mathbb{Z}_2$ (see e.g.~\cite{SRyu2010})
for the general number of $R$ replicas ensures that point-like topological defects (``vortices'') are possible. Here $G/H$ denotes the target manifold of the associated sigma model for each class  \cite{Zirnbauer96,Schnyder08,Evers08}.
}.
(The symplectic localization problem is described by a replicated sigma model with target manifold given by the orthogonal 
Grassmannian \cite{Zirnbauer96,Schnyder08,Evers08}
rather than the orthogonal group studied here.)

Let us apply this logic to the MIPT problem. For $R = 2$,
the field variable $\hat{X}$ is an abelian rotation matrix
parameterized by a single compact scalar field $\phi \in [0,2\pi)$,
and Eq.~(\ref{S}) reduces to
\begin{align}\label{SAB}
    S
    =
    \int d t \, d x
    \left\{
        \frac{K}{2}
        \Nabla \phi \cdot \Nabla \phi
        -
        \frac{M}{32}
        \,
        \cos(4 \phi)
    \right\}.
\end{align}
Here the stiffness $K = \lambda/4$.
The mass $M$ couples to a combination of vertex operators that share the same scaling dimension
$\Delta_M = 4/\pi K$.
A simple spacetime vortex in $\phi(t,x)$ would be induced by the dual 
(``magnetic'' \cite{BYB})
vertex operator $e^{i 2 \pi K \theta}$, 
which carries dimension 
$\Delta_V = \pi K$.
At the Kosterlitz-Thouless transition $K = K_c = 2/\pi$,
$\Delta_M = \Delta_V = 2$, meaning that both operators are marginal 
perturbations. 

In what follows, it is useful to view the above as
the $R \rightarrow 2$ limit of a \emph{non-abelian bosonization} scheme.
Specifically, we consider the Wess-Zumino-Novikov-Witten (WZNW) model
SO($R$)$_q$, where $q$ denotes the level 
\cite{EWitten1984,VGKnizhnik1984,BYB}.
The WZNW theory takes the same form as the original action in Eq.~(\ref{S}), 
augmented by a WZNW term.
We emphasize however that this is just a technical trick
(dualization) 
that we employ to encode the $R = 2$ physics. 
The effective field theory Eq.~(\ref{S}) derived directly from monitored Majorana fermions 
for generic $R$ does not possess a WZNW term.
Our logic is to capture the free-boson plus
mass and vortex perturbations in an equivalent, but more flexible CFT framework. 

The physics of the $R = 2$ model can be described by SO($R$)$_q$ with
level $q = 8$, see Eqs.~(\ref{KTKT})--(\ref{DeltaD}) and the surrounding text.
Let $y_K\equiv K_c - K$ denote the deformation of the stiffness away 
from $K_c$, while dimensionless couplings for the mass and vortex
fugacity are respectively denoted $\{y_M,y_V\}$.
Using the SO($R$)$_q$ encoding, we deform to $R = 2 - \epsilon  < 2$ to get the
lowest-order RG equations
\begin{align}\label{NABRGe}
\begin{aligned}
    \frac{d y_K}{d \ln L}
    =&\,
    \frac{\epsilon}{4}
    \,
    y_K
    +
    2\pi\left(y_V^2 - y_M^2\right),
\\
    \frac{d y_M}{d \ln L}
    =&\,
    \frac{\epsilon}{4}
    \,
    (1-x)
    \,
    y_M
    -
    4\pi
    \,
    y_K
    \,
    y_M,
\\
    \frac{d y_V}{d \ln L}
    =&\,
    \frac{\epsilon}{4}
    \,
    (4-x)
    \,
    y_V
    +
    4\pi
    \,
    y_K
    \,
    y_V.
\end{aligned}
\end{align}
Here $x$ denotes a real parameter that determines
the \emph{level deformation} of the WZNW model: $q \equiv 8 - x\, \epsilon$. 
Setting $\epsilon = 0$ recovers the KT-like physics of the $R = 2$ case,
with competing mass- and vortex-driven instabilities into the
volume- and area-law phases, respectively. 

Independent of the level deformation parameter $x$, 
the stiffness coupling $y_K$ acquires the positive dimension $\epsilon/4$.
Ignoring the mass and vortex couplings, the WZNW critical point 
is itself unstable to the deformation of the stiffness. 
This is qualitatively the same ``antilocalizing effect'' 
seen in the weak-coupling beta function for $\lambda$,
because $y_K \rightarrow -\infty$ corresponds to the flow
$\lambda \rightarrow \infty$ [Eq.~(\ref{RGpert})].
The latter flow results in a 
``thermal metal'' phase for monitored free Majorana fermions,
with entanglement scaling logarithmically enhanced beyond the area-law.

\begin{figure}[t!]
\centering
\includegraphics[width=0.5\textwidth]{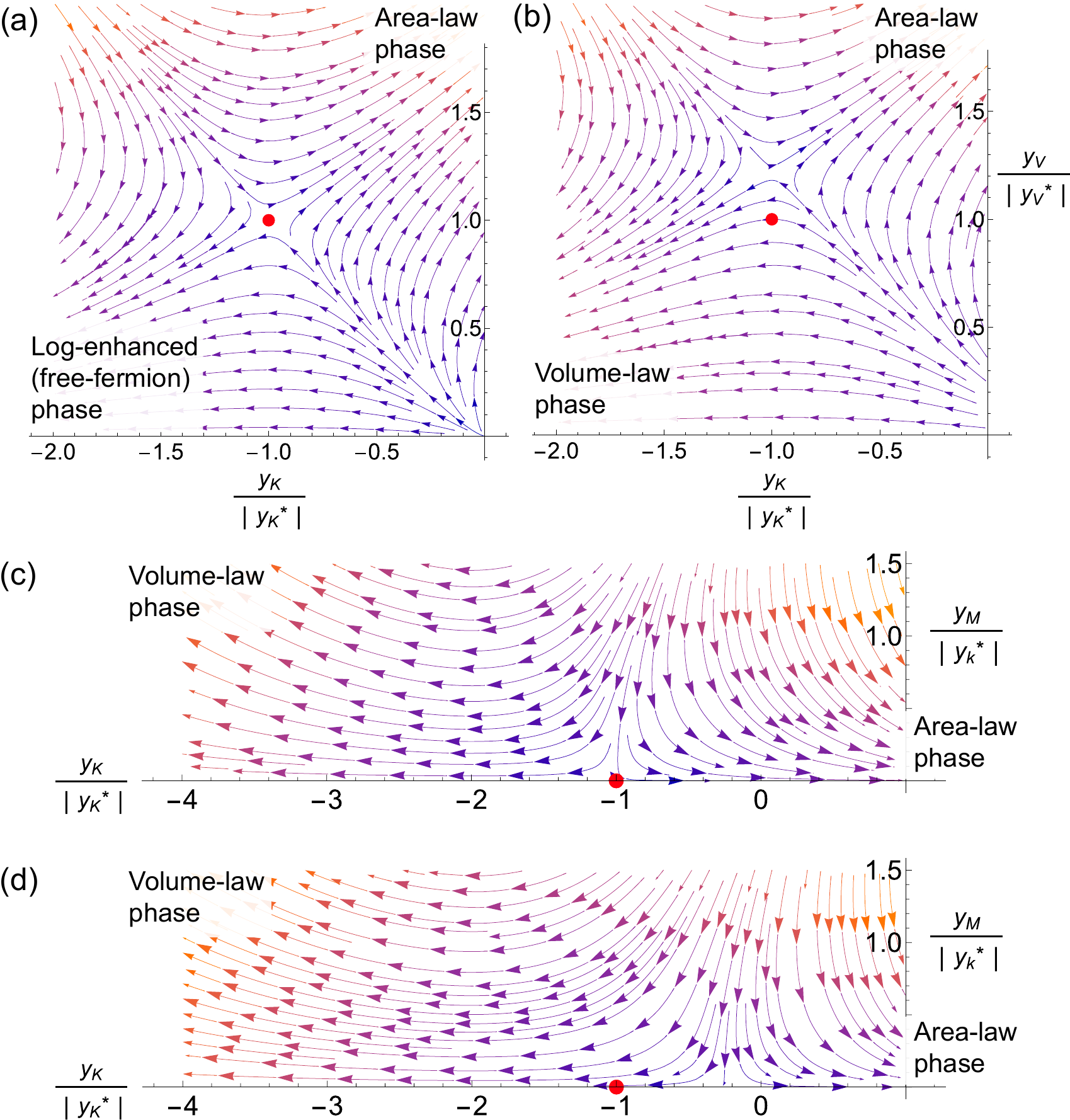}
\caption{(a,b) 
RG flows in the negative stiffness deviation $y_K$, vortex fugacity $y_V$ plane,
from Eq.~(\ref{NABRGe}). Here we set $\epsilon = 1$ and $x = 3$ (see text).
Subpanel (a) shows the flow in the noninteracting plane $y_M = 0$; the
noninteracting MIPT [Eq.~(\ref{yNI})] is indicated by the red dot.
The mass parameter $y_M$ is an irrelevant perturbation to this fixed point.
Subpanel (b) 
depicts 
the flow in the plane $y_M = |y_K^*| > 0$; the red dot still labels
the MIPT at $y_M = 0$.  
The mass is a \emph{dangerously irrelevant} perturbation,
because flow through the dot now runs towards $y_K \rightarrow -\infty$ (large stiffness)
and $y_V \rightarrow 0$ (vanishing fugacity).  
Panels (c) and (d) show that this in turn drives a flow towards large 
mass $y_M$, signaling the onset of the volume-law phase. 
(c,d) depict the RG flow in the stiffness-mass ($y_K,y_M$) plane, 
with the fugacity set equal to its critical value $y_V^*$ and $y_V^*$/2, respectively.
For any nonzero $y_V$, the transition between volume- and area-law phases is tuned by
the stiffness $y_K$ alone, as also indicated by (a,b). The mass flows to zero
in the area-law phase and along the critical inflow trajectory.
By contrast, tuning past the critical manifold ultimately produces a 
rapidly growing mass. The latter is consistent with
the stabilization of the volume-law phase by a finite Fermi's golden rule inelastic scattering rate density, which counters the disentangling effects of the measurements.
}
\label{Fig--Flow}
\end{figure}

Even for free fermions, however, we should not ignore the effect of the
vortex fugacity $y_V$ near the putative MIPT critical point.
Eq.~(\ref{NABRGe}) possesses two nontrivial fixed points: a multicritical interacting 
one with $y_M \neq 0$, and a \emph{noninteracting} one [Eq.~(\ref{yNI})].
Linearizing the flow around 
the latter
fixed point
gives the RG eigenvalues 
$
\left\{
            \left(1 \pm \sqrt{33 - 8 \, x}\right),
        2(5 - 2 \, x)
    \right\}
    (\epsilon/8).
$
This fixed point
possesses a single
relevant direction for $5/2 < x < 4$. 
Because the noninteracting MIPT is driven by defect proliferation,
the vortex fugacity $y_V$ should carry a positive (relevant) scaling
dimension. Eq.~(\ref{NABRGe}) then implies that $x < 4$.
We conjecture that 
$x > 5/2$,
so that Eq.~(\ref{yNI}) locates 
the MIPT for \emph{both} the interacting and noninteracting
monitored fermions. 
In this case, the single relevant direction sets the correlation length exponent 
$
    \nu = (1 + \sqrt{33 - 8\, x})^{-1}(8/\epsilon).
$
This satisfies the Chayes-Harris criterion
\cite{Chayes86,Harris74} for all $\epsilon \leq 1$.
Large-scale numerical results on monitored, free-fermion dynamics identified a noninteracting MIPT with $\nu \sim 2.1$ \cite{Jian23}, which would 
(i.e.\ using the 1-loop values) correspond to 
$x = 3$.

The noninteracting MIPT separates area-law and logarithmic ``thermal metal'' phases
for monitored, noninteracting Majorana fermions. 
At our candidate fixed point in Eq.~(\ref{yNI}) with 
$5/2 < x < 4$, 
the mass operator
is \emph{dangerously} irrelevant. I.e., setting the stiffness and fugacity couplings
to their fixed-point values $y_{K,V} = y_{K,V}^*$, a deviation $y_M > 0$ induces an instability
that ultimately flows into the volume-law phase with $y_M \rightarrow \infty$.
This is illustrated in Fig.~\ref{Fig--Flow}.
By contrast, tuning the fugacity and/or stiffness away from their fixed-point values
allows access to the area-law phase, where instead $y_K \rightarrow \infty$ and the mass remains irrelevant.


\section{Technical approach \label{sec:details}}

\subsection{Microscopic model \label{sec:micro}}

The noninteracting part of the model consists 
of spinless fermions that form a 1D mean-field $p$-wave superconductor,
with the Hamiltonian
\begin{align}\label{H0}
	H_0 
	=&\, 
	\sum_{x}
	\left[
		-J 
		\, 
		c_x^\dagger\left(c_{x+a} + c_{x - a}\right)
		+
		(2 J - \mu)
        \,
		c_x^\dagger \, c_x
	\right]
\nonumber\\
&\,
	-
	\frac{\ii \, \Delta}{4}
	\sum_{x}
	c_x
	\left(c_{x+a} - c_{x - a}\right)
	+
    \textrm{H.c.},
\end{align}
where $a$ is the lattice constant and
H.c.\ denotes the Hermitian conjugate.
The model is diagonalized by a Bogoliubov transformation,
giving the quasiparticle dispersion 
\begin{align}\label{EkDef}
	E(k) = \sqrt{\tilde{\varepsilon}^2(k) + \Delta^2 \sin^2(k a)},
\end{align}
where $\tilde{\varepsilon}(k) = 2J[1 - \cos(ka)] - \mu$.
Without monitoring, the model is in class BDI, with
a gapped topological ground state in the weakly-paired regime
with $\Delta \neq 0$ and $0 < \mu < 4J$.

Eq.~(\ref{H0}) can be alternatively cast in terms of two species
of lattice Majorana fermions,
\begin{align}
	\gamma_x \equiv \frac{1}{\sqrt{2}}\left(c_x + c^\dagger_x\right),
\qquad
	\xi_x \equiv \frac{\ii}{\sqrt{2}}\left(c_x - c^\dagger_x\right),
\end{align}
leading to
\begin{align}\label{H02Maj2}
	H_0 
	=&\,
	\ii
	\sum_{x}
\left[
	\frac{\Delta}{2} 
	\,
	\Big(
		\xi_x \, \xi_{x+a}
		-
		\gamma_x \, \gamma_{x+a}
	\Big)
	+
	(2 J - \mu)
	\,
	\xi_{x} \, \gamma_{x}
\right]
\nonumber\\
&\,
    +
	\ii \, J
	\sum_{x}
	\Big(
		\gamma_{x} \, \xi_{x+a}
	-
		\xi_{x} \, \gamma_{x+a} 
	\Big).
\end{align}
The hopping term $J$ results in a \emph{dimerization} of the Majorana fields. 

When formulated as a Keldysh path integral subject to weak monitoring of the fermion density,
it can be shown that the noninteracting Keldysh action possesses SO($R$)$\times$SO($R$) symmetry, 
associated to independent replica rotations on the forward and backward Keldysh contours. 
This is class DIII. A fine-tuned version of the model with $J = \mu = 0$, so that the Hamiltonian in Eq.~(\ref{H02Maj2}) decomposes into two decoupled Majorana chains, possesses a larger symmetry in this formulation. In this case the monitored Keldysh action exhibits SO($2R$) symmetry, wherein replica rotations between the contours are symmetries. This corresponds to class D. 
See 
Appendix~\ref{app:sym}
for details.

\subsection{Numerical tests---additional details \label{sec:numdetails}}

The fact that average value of $G$, defined to be the square of the fermion-fermion correlation function at the final time-slice of
the circuit, must carry dimension one for the noninteracting monitored circuit can be understood as follows.  
This is clearly so at the weakly coupled (Gaussian) fixed point of the SO$(R)$ principal chiral model (PCM) in \cite{Jian23,MFava2023} at which the SO$(R)$ target manifold has 
zero
curvature. As the coupling strength of the PCM is increased, the thereby introduced finite curvature of the manifold renormalizes this fermion bilinear operator in perturbation theory. However, global 
SO$(R)$$\times$SO$(R)$ symmetry implies that there will be two Noether currents due to right- and left- multiplication symmetry by SO$(R)$, which must have scaling dimensions of exactly unity. At the circuit boundary at the final time-slice only one linear combination will survive due to the ``absorbing boundary condition''~\cite{CMJian2022} that prevails there.
This means that the quantum  trajectory average of the square of the fermion correlation function $G$ at the final time-slice decays, at the MIPT, 
precisely~\footnote{This argument was presented in a talk (by AWWL) at the Simons Center for Geometry and Physics in August 2023. The same argument was also presented later 
in~\cite{Mirlin2024_Above1D}.}
with exponent $2 = 2\times1$.
This exponent is seen numerically in~\cite{Jian23}.\\

\subsection{Bosonization and epsilon expansion \label{sec:bosonization}}

We consider first the case of two replicas, $R = 2$ and Eq.~(\ref{SAB}).
At the Kosterlitz-Thouless transition $K = K_c = 2/\pi$,
$\Delta_M = \Delta_V = 2$, meaning that both the mass and vortex operators are marginal 
perturbations. Let $y_K\equiv K_c - K$ denote the deformation of the stiffness away 
from $K_c$, while dimensionless couplings for the mass and vortex
fugacity are respectively denoted $\{y_M,y_V\}$, as in Eq.~(\ref{NABRGe}). 
Using standard techniques \cite{Cardy,Giamarchi},
the one-loop RG equations are 
(Appendix~\ref{app:R=2})
\begin{align}\label{KTKT}
\begin{gathered}
    \frac{d y_K}{d \ln L} = 2\pi \left(y_V^2 - y_M^2\right),
\quad
    \frac{d y_M}{d \ln L} = - 4 \pi \, y_K \, y_M,
\\
    \frac{d y_V}{d \ln L} = 4 \pi \, y_K \, y_V.
\end{gathered}
\end{align}
These equations represent the competing 
instabilities due to mass and vortex perturbations.
For $y_K = 0$ there are critical fixed lines $y_V = \pm y_M$. 
Tuning $|y_V| > |y_M|$ sends $y_K \rightarrow + \infty$, corresponding
to vortex proliferation; the mass coupling $y_M$ flows quickly to zero.
Tuning $|y_V| < |y_M|$ sends $y_K \rightarrow - \infty$, 
with explicit O($2$) symmetry breaking due to the mass term
and vanishing vortex fugacity $y_V \rightarrow 0$.
Although $R = 2$ is a toy version, we can loosely associate
these two instabilities to the area- and volume-law phases,
respectively.

Eq.~(\ref{KTKT}) is obtained via abelian bosonization techniques 
in Appendix~\ref{app:R=2}.
In the SO($R$)$_q$ non-abelian bosonization scheme
detailed in Appendix~\ref{app:NAB},
we identify the mass operator in Eqs.~(\ref{S}) and (\ref{SAB}) as a diagonal primary field associated to 4th-rank, traceless and completely symmetrized tensors. 
The scaling dimension is 
\cite{VGKnizhnik1984},
\begin{align}\label{DeltaM}
    \Delta_M
    =
    \frac{4(R+2)}{D_q(R)},
\end{align}
where
\begin{align}\label{WZNWDen}
    D_q(R)
    \equiv  
    R - 2 + q.
\end{align}
The bosonization of the vortex operator
should involve spinor (``twist'') representations of SO($R$)$_q$. 
Completely symmetrized tensors built from the (e.g.) right-handed spinor representation 
of SO($R$)$_q$ reduce to single-component
primary fields in the $R \rightarrow 2$ limit. 
For a rank-$p$ tensor, the conformal dimension scales like $p^2$ in this limit, as expected for a family of vertex operators
(see Appendix \ref{app:NAB} for details).
This allows us to identify the vortex operator with the choice $p = 8$, leading to 
\begin{align}\label{DeltaD}
    \Delta_V = \frac{R(R+6)}{D_q(R)}.
\end{align}
For the choice of level $q = 8$,
$\Delta_V = \Delta_M = 2$ (as desired) 
and moreover the WZNW action 
perturbed by the mass operator
precisely reproduces 
Eq.~(\ref{SAB}) at $K = K_c$, in the limit $R \rightarrow 2$.

We are supposed to take the $R \rightarrow 1$ limit in order to restore the Born rule and access the MIPT \cite{Jian20,Bao20}.
We will use the SO($R$)$_q$ framework in order to deform the theory away from $R = 2$. This is similar in spirit but different in details to the Fu-Kane calculation \cite{FuKane12}. 

The critical fixed lines in Eq.~(\ref{KTKT}) are unlikely to survive the continuation to $R < 2$.
Indeed, the operator dimensions $\Delta_{M,V}$ in SO($R$)$_q$ deform away from 2, while the stiffness coupling perturbation $y_K$ acquires a nonzero dimension. Incorporating these dimensions into Eq.~(\ref{KTKT}), we get 
\begin{subequations}\label{NABRG}
\begin{align}
    \frac{d y_K}{d \ln L}
    =&\,
    \frac{2(2-R)}{D_q(R)}
    \,
    y_K
    +
    2\pi\left(y_V^2 - y_M^2\right),
\label{ykRG}
\\
    \frac{d y_M}{d \ln L}
    =&\,
    \left[
        2
        -
        \frac{4(R+2)}{D_q(R)}
    \right]
    y_M
    -
    4\pi
    \,
    y_K
    \,
    y_M,
\\
    \frac{d y_V}{d \ln L}
    =&\,
    \left[
        2
        -
        \frac{R(R+6)}{D_q(R)}
    \right]
    y_V
    +
    4\pi
    \,
    y_K
    \,
    y_V.
\end{align}
\end{subequations}
Setting $R = 2 - \epsilon$, $q = 8 - x \, \epsilon$, 
and expanding to lowest nontrivial order in $\epsilon$ gives
Eq.~(\ref{NABRGe}).
In those equations, we neglect $\ord{\epsilon}$ corrections
to the OPE coefficients (quadratic terms). These do 
not affect the results to lowest nontrivial order in $\epsilon$.

The noninteracting fixed point of Eq.~(\ref{NABRGe}) is given by 
\begin{align}\label{yNI}
    \{y_K^*,y_M^*,y_V^*\}
    =
    \left\{
    (x - 4),
    0,
    \sqrt{2(4 - x)}
    \right\}
    (\epsilon / 16\pi),
\end{align}
which exists for $x \leq 4$.


\section{Conclusion \label{sec:conclusion}}

In this paper, we considered the MIPT for the generic problem 
of monitored, interacting fermions (equivalent to a qubit chain with conserved fermion parity). 
We argued that the continuous vanishing of the entangling rate density implies the vanishing
(dangerous irrelevance) of a mass term in the effective field theory. In the absence
of conserved quantities, this mass term is the \emph{only} imprint of the interactions in the field theory, and takes the form of a replica-sector ``jump term.'' 

In the considered symmetry class, a vanishing mass implies that the MIPT for interacting fermions must be the same as the \emph{noninteracting} one. The latter arises for monitored, noninteracting Majorana fermions, where it separates the area-law phase at strong monitoring from a logarithmically-enhanced area-law phase (``thermal metal'') at weak monitoring. We enumerated a list of numerical tests for this idea. In addition, we constructed an analytically controlled field theory of the noninteracting MIPT,
and showed how the mass can be irrelevant there. 

Key directions for future research include the following.
\begin{enumerate}
\item{The numerical tests summarized in Sec.~\ref{sec:num} can be used to evaluate the equivalence of the noninteracing and interacting MIPT fixed points. Note that this program can be carried out independent of the analytical theory developed here.}
\item{In the analytical theory of the MIPT presented in Sec.~\ref{sec:analytical}, the level deformation parameter $x$ is not determined. It is found that the value $x \simeq 3$ gives a phenomenology consistent with the picture articulated here (irrelevant mass operator) and existing numerical results $\nu \simeq 2.1$ \cite{Jian23}. Is there an additional analytical input or constraint that fixes (or further bounds) the value of $x$? 
Such a constraint might play the role of a ``non-unitary $c$-theorem'' for deformed WZNW levels. 
}
\item{In sharp constrast to the case considered here, monitored fermions with conserved U(1) charge (class AIII) possess only the area-law in the absence of interactions \cite{Poboiko23,Fava24}. With interactions, these are described by an SU($R$) replica sigma model effective field theory that incorporates additional Noether current-current interactions \cite{Guo24}. If the same phenomenology conjectured here applies in this case, we expect that the mass term is also dangerously irrelevant at the interacting class-AIII MIPT fixed point. The latter should then be determined by the \emph{replica-anisotropic} SU($R$) sigma model deformed by the Noether current-current perturbations (which reduce the symmetry to $S_R$ $\times$ $S_R$). This fixed point must be different from the putative interacting class-DIII one that should exhibit emergent SO($R$) $\times$ SO($R$) symmetry according to the arguments in this paper.
}
\end{enumerate}


\acknowledgements
This work was supported by the Alfred P. Sloan Foundation through a Sloan Research Fellowship (C.-M.J.). One of us (A.W.W.L.) is grateful to the Kavli Institute for Theoretical Physics (KITP), which is supported by the National Science Foundation by grant NSF PHY-2309135, and the KITP Programs ``Noise-robust Phases of Quantum Matter'' and ``Learning the Fine Structure of Quantum Dynamics in Programmable Quantum Matter,'' where part of this work was carried out.

\bibliography{Refs}

\appendix

\section{From the microscopic model to the field theory  \label{app:sym}}

The noninteracting part of the microscopic Hamiltonian was given by 
Eqs.~(\ref{H0}) and (\ref{H02Maj2}). Incorporating density-density interactions and weak monitoring of the fermion density,
the theory can be formulated as a replicated Keldysh path integral, 
\begin{align}
    Z
    =
    \int \, \mathcal{D}\bar{c} \, \mathcal{D} c \, \mathcal{D} a \, e^{i S},
\end{align}
\begin{align}
	S
	=&\,
	S_0
	+
	S_n
	+
	S_a,
\end{align}
where
\bsub
\begin{widetext}
\begin{align}
	S_0
	\equiv&\,
	\sum_{\tau = \pm}
	(\tauh^3)_{\tau \tau}
    \int_{\bz}
    \frac{d k}{2 \pi}
	\int
    d t
    \,
	\bigg\{
		\bar{c}_{\tau,j}(k) 
		\Big[
			\ii \, \parr_t  - \te(k)
		\Big] 
		c_{\tau,j}(k)
		-
		\frac{\Delta}{2} \, \sin(k)
		\Big[
			\bar{c}_{\tau,j}(k) \, \bar{c}_{\tau,j}(-k) 
			+  
			c_{\tau,j}(-k) \,  c_{\tau,j}(k)
		\Big]
	\bigg\},
\\	
	S_n
	\equiv&\,
	-
	\sum_{x}
	\int 
    d t
	\Bigg[
		n_{\cl,x,j} \, a_{\q,x,j}
		+
		n_{\q,x,j} \, a_{\cl,x,j}
		-
		\ii
		\,
		\upsilon(t,x)
		\,
		\sum_{j}
		n_{\cl,x,j}
	\Bigg],
\end{align}
\end{widetext}
\begin{align}
	S_a
	\equiv&\,
	\frac{2}{U}
	\sum_{x}
	\intl{t}
	a_{\cl,x,j} \,\, a_{\q,x,j}.
\label{SM--SaDef}
\end{align}
\esub
Here $\tau \in \{T,\bar{T}\}$ is the Keldysh-contour index,
$\tauh^3$ is a Keldysh-space Pauli matrix, 
and $j \in \{1,2,\ldots,R\}$ counts replicas (doubly repeated replica indices are summed). 
The pairing is taken to be purely ``classical'' $\Delta = \Delta_\cl$, i.e.\ static and homogeneous \cite{Liao17}.
The boson field $a_{\cl,\q}$ mediates instantaneous, short-ranged density-density interactions of strength $U$;
the fermion classical and quantum densities are (suppressing Keldysh indices)
\begin{align}
    n_{\cl,x,j} \equiv \bar{c}_{x,j} \, c_{x,j},
\qquad
    n_{\q,x,j} \equiv \bar{c}_{x,j} \, \tauh^3 \, c_{x,j}.
\end{align}
Weak measurements of the density are encoded in the ``measurement noise'' $\upsilon(t,x)$ \cite{Guo24}.
We transform to the
Larkin–Ovchinnikov (LO)
basis \cite{Kamenev23} via
\begin{gather}
\begin{gathered}
	c \rightarrow \tauh^3 \, U_{\mathsf{LO}} \, c,
\qquad
	\bar{c} \rightarrow \bar{c} \, U_{\mathsf{LO}}^\dagger,
\\
	U_{\mathsf{LO}} = U_{\mathsf{LO}}^* = \frac{1}{\sqrt{2}}\left(\hat{1} + \ii \, \tauh^2\right),
\end{gathered}
\end{gather}
which leads to 
\begin{align}
    \!
	S_0
	\rightarrow&\,
    \int_{\bz}
    \frac{d k}{2 \pi}
	\int
    d t
    \,
	\bigg\{
		\bar{c}_{j}(k) 
		\Big[
			\ii \, \parr_t  - \te(k)
		\Big] 
		c_{j}(k)
	\bigg\}
\nonumber\\
    &\,
    -
    \int_{\bz}
    \frac{d k}{2 \pi}
	\int
    d t
    \,
	\bigg\{
		\frac{\Delta}{2} \, \sin(k) \,
			\bar{c}_{j}(k) \, \tauh^1 \, \bar{c}_{j}^\T(-k) 
	\bigg\}
\nonumber\\
    &\,
    -
    \int_{\bz}
    \frac{d k}{2 \pi}
	\int
    d t
    \,
	\bigg\{
		\frac{\Delta}{2} \, \sin(k) \,
			c_{j}^\T(-k) \, \tauh^1 \,  c_{j}(k)
	\bigg\}.\!\!
\end{align}
We define a two-component real spinor \cite{Liao17}
\begin{align}\label{SM--chiDef}
\begin{aligned}
	\chi(t,k)
	\equiv&\,
	\begin{bmatrix}
		c(t,k)
	\\
		\tauh^1 \, \bar{c}^\T(t,-k)
	\end{bmatrix}_\sigma,
\\
	\bar{\chi}(t,k)
	\equiv&\,
	\chi^\T(t,-k) \, \tauh^1 \, \Mp
\\
	=&\,
	\begin{bmatrix}
		\bar{c}(t,k)
	&
		{c}^\T(t,-k) \, \tauh^1
	\end{bmatrix}_\sigma.
\end{aligned}
\end{align}
Here $\sigma$ denotes particle-hole space, and 
\begin{align}
    \Mp \equiv \sigh^1
\end{align}
is a particle-hole matrix.
Then
\begin{align}\label{SM--S0F}
	S_0	
	=&\,
	\frac{1}{2}
    \int_{\bz}
    \frac{d k}{2 \pi}
	\int
    d t
    \,\,
	\bar{\chi}_j(t,k)
	\left[
		\ii \, \parr_t
		-
		\hat{h}(k)
	\right]
	\chi_j(t,k)
\nonumber\\
    =&\,
	\frac{1}{2}
	\sum_{x,x'}
	\int
    d t 
    \,\,
	\bar{\chi}_{x,j}(t)
	\left[
		\ii \, \parr_t
		\,
		\delta_{x,x'}
		-
		\hat{h}_{x,x'}
	\right]
	\chi_{x',j}(t).
\end{align}
Here the Bogoliubov-de Gennes Hamiltonian is
\begin{align}\label{SM--BdG}
    \hat{h}(k)
    \equiv
		\begin{bmatrix}
			\te(k)
		&
			\Delta \, \sin(k)
		\\
			\Delta \, \sin(k)
		&
			-\te(k)
		\end{bmatrix}_\sigma,
\end{align}
which satisfies the particle-hole condition
\begin{align}
    - \Mp \, \hat{h}^\T(-k) \, \Mp = \hat{h}(k).
\end{align}
We also have
\begin{align}
	n_{\cl,j} = \frac{1}{2} \, \bar{\chi}_j \, \sigh^3 \, \tauh^1 \, \chi_j,
\quad
	n_{\q,j} = \frac{1}{2} \, \bar{\chi}_j \, \sigh^3 \, \chi_j,
\end{align}
and
\begin{align}\label{SM--SnF}
	S_n
	=&\,
	\frac{\ii}{2}
	\sum_{j = 1}^R
	\sum_{x}
	\int 
    d t
	\,
	\upsilon(t,x)
    \,
	\bar{\chi}_{x,j}(t)
	\,
	\sigh^3
	\,
	\tauh^1
    \,
	\chi_{x,j}(t)
\nonumber\\
&\,
	-
	\frac{1}{2}
	\sum_{j = 1}^R
	\sum_{x}
	\int 
    d t
    \,\,
	\bar{\chi}_{x,j}
	\,
	\sigh^3
	\left[
		a_{\q,x,j}
		\,
		\tauh^1 
		+
		a_{\cl,x,j}
	\right]
	\chi_{x,j}.
\end{align}

\subsection{Symmetry classification (target manifold)}

The target manifold of the replicated sigma model for the problem of \emph{noninteracting} monitored fermion dynamics can be determined by a symmetry analysis of the system in an arbitrary, fixed quantum trajectory $\upsilon(t,x)$ \cite{Guo24}. We consider a maximal unitary transformation of the fermion field $\chi$ in the combined 
particle-hole ($\sigma$) $\otimes$ LO ($\tau$) $\otimes$ replica spaces,
\begin{align}
    \chi \rightarrow \hat{U} \, \chi,
    \qquad 
    \bar{\chi} \rightarrow \bar{\chi} \, \sigh^1 \, \tauh^1 \, \hat{U}^\T \, \sigh^1 \, \tauh^1.
\end{align}
Here $\hat{U}$ is a U($4R$) matrix.

We focus first upon the \emph{fine-tuned model} with $J = \mu = 0$, which corresponds to 
two decoupled Majorana chains without monitoring [see Eq.~(\ref{H02Maj2})].  
Then the BdG Hamiltonian in Eq.~(\ref{SM--BdG}) is purely anomalous, $\hat{h} \propto \sigh^1$. 
In this special case, invariance of Eqs.~(\ref{SM--S0F}) and (\ref{SM--SnF}) with $a_{\cl} = a_{\q} = 0$ (i.e., turning off the interactions) imposes three constraints,
\begin{align}\label{SM--Constraints}
\begin{aligned}
    \hat{U}^\T \sigh^1 \, \tauh^1 \hat{U} =&\, \sigh^1 \, \tauh^1,
\\
    \hat{U}^\dagger \sigh^3 \, \tauh^1 \hat{U} =&\, \sigh^3 \, \tauh^1,
\\
    \hat{U}^\dagger \sigh^1 \hat{U} =&\, \sigh^1.
\end{aligned}
\end{align}
These constraints restrict $\hat{U} \in$ SO(2$R$). 

To derive the sigma model, we 
average over measurement trajectories, 
decouple the resulting four-fermion term with a replica matrix field, 
and integrate-out the fermions. 
Calculating the spacetime homogeneous saddle-point configuration of the matrix field introduces a non-equilibrium fermion decay rate term into the effective fermion action \cite{Poboiko23,Guo24}. In this case, the saddle-point corresponds to a fermion bilinear
\begin{align}\label{SM--SP}
    \ii \, \gamma \, \bar{\chi} \, \tauh^3 \, \chi,
\end{align}
where $\gamma$ is the bare measurement rate. This respects the causal structure of the Keldysh theory (but not the fluctuation-dissipation theorem) in the LO basis \cite{Guo24}. 

The target manifold for the noninteracting theory is the set of fluctuations satisfying the constraints in Eq.~(\ref{SM--Constraints}) that produce a nontrivial rotation of the saddle-point in Eq.~(\ref{SM--SP}). This is the space SO($2 R$)/U($R$), the sigma-model manifold associated to class D \cite{Schnyder08}.

For the general model, Eqs.~(\ref{H0}) and (\ref{H02Maj2}) with nonzero $J$ and/or $\mu$, we need to impose the additional constraint 
\begin{align}
    \hat{U}^\dagger \sigh^3 \hat{U} = \sigh^3.
\end{align}
Then $\hat{U}$ cannot depend on particle-hole space, and $\hat{U} \in$ SO($R$)$\times$SO($R$).
Modding out fluctuations that preserve the saddle-point [Eq.~(\ref{SM--SP})] gives 
\begin{align}\label{SM--DIII}
    \frac{\textrm{SO}(R)\times\textrm{SO}(R)}{\textrm{SO}(R)}
    \sim
    \textrm{SO}(R),
\end{align}
corresponding to the class-DIII sigma model \cite{Schnyder08}.

\subsection{Comparison to U(1)-symmetric case}

As an aside, we note that  
the monitored dynamics of charge-conserving free fermions
can be obtained by taking $\Delta = 0$ in Eq.~(\ref{SM--BdG}). 
Then the last constraint in Eq.~(\ref{SM--Constraints}) is replaced
with 
\begin{align}\label{SM--Constraints-U(1)}
    \hat{U}^\dagger \sigh^3 \hat{U} =&\, \sigh^3.
\end{align}
The resulting $\hat{U}$ can be parameterized in terms of 
two independent unitary matrices, giving 
$\textrm{U}(R)\times\textrm{U}(R)$ symmetry.
In this case, there is only an area-law phase 
without interactions
and no MIPT
in 1+1-dimensions \cite{Poboiko23,Fava24}.

\subsection{Gradient expansion}

To derive the effective field theory, one averages over measurement trajectories and
decouples the resulting four-fermion term with a matrix Hubbard-Stratonovich field. Following the same steps outlined in Ref.~\cite{Guo24}, one arrives at the effective Gaussian action for the 
replicated entanglement field theory,
\begin{align}\label{SM--SGauss}
    S
    \simeq
    \frac{\lambda}{8}
    \int d t \, d x \, 
    \tr\left[
    \begin{aligned}
    &\,
        \parr_t \hat{W}^\T \parr_t \hat{W}
        +
        v^2
        \,
        \parr_x \hat{W}^\T \parr_x \hat{W}
    \\&\,
    +
        m^2
        \,
        \hat{W}^\T \hat{W}
    \end{aligned}
    \right],
\end{align}
where $\hat{W}^\T = - \hat{W} \rightarrow W_{j,k}$ is an antisymmetric, real,
but otherwise unconstrained replica matrix field.
[The LO Keldysh ($\tau$) and particle-hole ($\sigma$) spaces have been traced out in the trace-log expansion.]
In Eq.~(\ref{SM--SGauss}), 
$v = v(J,\Delta,\mu)$ is the 
entanglement-sector
``velocity,''
and $m^2 \equiv M/\lambda \propto \lambda \, U^2$ is the mass 
(determined by a Fermi's golden-rule like fermion self-energy \cite{Guo24}). 
The stiffness $\lambda$ is inversely proportional to the measurement rate. 
The nonlinear theory in Eq.~(\ref{S}) obtains by setting the velocity to one and 
promoting the fluctuation
\[
    \hat{W} 
    \rightarrow 
    \hat{W} + \sqrt{\hat{1} + \hat{W}^2} - \hat{1}
    \equiv
    \hat{X} - \hat{1},
\]
so that $\hat{X}^\T \, \hat{X} = \hat{1}$.
This constrains $\hat{X}$ to the Goldstone manifold specified by Eq.~(\ref{SM--DIII}).

\section{$R = 2$ operator content and 1-loop RG \label{app:R=2}}

For a compact boson field $\phi = \phi + 2 \pi$, defined over a finite 1D spatial volume $L$ with winding boundary condition 
$\phi(t,x + L) = \phi(t,x) - 2 \pi m$, $m \in \mathbb{Z}$, the set of primary fields is quantized and can be expressed as \cite{BYB}
\begin{align}\label{SM--VertOp}
    V_{n,m} = :e^{\ii \left(n \, \phi + 2 \pi K  m \, \theta\right)}:,
\quad
    n \in \mathbb{Z},
\end{align}
where $K$ is the stiffness and $\phi$ is the field in Eq.~(\ref{SAB}), 
and $\theta$ is the dual (axial) field.
The operator product expansion (OPE) between two such operators with opposite polar and axial U(1) charges is
\begin{align}\label{SM--OPE} 
    V_{n,m}&(z,\bar{z}) \, V_{-n,-m}(w,\bar{w})
\nonumber\\
    \sim&\,
    \frac{1}{(z - w)^{2 h_{n,m}}(\bar{z} - \bar{w})^{2 \bar{h}_{n,m}}}
\nonumber\\
&\,
    +
    \frac{\alpha_{n,m} \, \bar{\alpha}_{n,m}
    }{(z - w)^{2 h_{n,m} - 1}(\bar{z} - \bar{w})^{2 \bar{h}_{n,m}-1}}
    J(w) \, \bar{J}(\bar{w})
\nonumber\\
&\,    +
    \ldots,
\end{align}
where
\begin{align}\label{SM--abeliandata}
\begin{aligned}
    \{\alpha_{n,m},\bar{\alpha}_{n,m}\}
    =&\,
    \left(n \pm 2 \pi K m\right)/\sqrt{4 \pi K},
\\
    \{h_{n,m},\bar{h}_{n,m}\}
    =&\,
    \left\{{\alpha_{n,m}^2}/{2},{\bar{\alpha}_{n,m}^2}/{2}\right\}.
\end{aligned}
\end{align}
In Eq.~(\ref{SM--OPE}), $\{z,\bar{z}\} \equiv x \pm \ii \, t$, and the current operator 
$J(z) \equiv \ii \sqrt{4 \pi K} \, \parr \phi$, with $\parr \equiv d/d z$.

The stiffness, mass, and vortex fugacity operators can be identified as 
\begin{align}\label{SM--Ops}
\begin{aligned}
    O_J \equiv&\, (J \bar{J}),
\\
    O_M \equiv&\, {\textstyle{\frac{1}{\sqrt{2}}}}(V_{4,0} + V_{-4,0}),
\\
    O_V \equiv&\, {\textstyle{\frac{1}{\sqrt{2}}}}(V_{0,1} + V_{0,-1}).
\end{aligned}
\end{align}
The nonvanishing OPE coefficients between these operators
are [via Eq.~(\ref{SM--OPE}) and the U(1) Ward identities]
\begin{align}\label{SM--OPECoeff}
\begin{gathered}
    C_{J M M} = C_{M J M} = C_{M M J} = \Delta_M,
\\
    C_{J V V} = C_{V J V} = C_{V V J} = - \Delta_V.
\end{gathered}
\end{align}
At criticality $\Delta_M = \Delta_V = 2$. 
Then the one-loop RG in Eq.~(\ref{KTKT}) 
is determined by Eq.~(\ref{SM--OPECoeff}) \cite{Cardy}.

\section{Non-abelian bosonization \label{app:NAB}}

The SO($R$)$_q$ WZNW model \cite{BYB} takes the same form
as the NLsM in Eq.~(\ref{S}) with $M = 0$, 
\begin{align}\label{SWZNW}
    S_{\mathsf{WZNW}} 
    = 
    \frac{\lambda}{16}
    \int d t \, d x
        \trr\left[\Nabla \hat{Q}^T \cdot \Nabla \hat{Q} \right]
    +
    q \, \Gamma_{\mathsf{WZNW}}.
\end{align}
Here $\hat{Q}$ denotes a $R$$\times$$R$ SO($R$)-valued matrix, 
and $\Gamma_{\mathsf{WZNW}}$ is the WZNW term.
At $\lambda = q / \pi$, the theory is conformally invariant. 
The beta function for a deviation $y_K \propto (q/\pi - \lambda)$
is given by 
Eq.~(\ref{ykRG}) with $y_V = y_M = 0$;
the stiffness deviation is \emph{relevant} for $R < 2$. 

Scaling dimensions of primary fields in a WZNW theory are determined
by the Casimir eigenvalues of the associated irreducible representations
of the group. The mass operator 
\begin{align}\label{SM--OMDef}
    O_M 
    \equiv 
    \sum_{j,k = 1}^R
    X_{j k}^4
\end{align}
is an eighth-rank tensor composed of two groups of 4 identical left- and right-
indices, respectively. In the WZNW theory, this can be viewed as a diagonal
primary field with a holomorphic half transforming in the 4th-rank, traceless
symmetric tensor representation of SO($R$): 
\begin{align}
    \sum_{j_1,k_1 = 1}^R
    X_{j_1,k_1}^4
    \rightarrow&\,
    \sum_{j_1 = 1}^R T_{(j_1 j_1 j_1 j_1)}(z)
    \sum_{k_1 = 1}^R \bar{T}_{(k_1 k_1 k_1 k_1)}(\bar{z}),
\nonumber\\
    &\,
    \sum_{j_1 = 1}^R T_{(j_1 j_1 j_2 j_3)} = 0. 
\end{align}
Here $(j_1 j_2 \cdots j_p)$ denotes the complete symmetrization of the indices $\{j_1,\ldots,j_p\}$,
and each individual index transforms in the vector representation of SO($R)$.
The corresponding
highest weight is $\Lambda = 4 \, \omega_1$, where $\omega_1$ is the first fundamental weight. 
This determines $\Delta_M$ in Eq.~(\ref{DeltaM}). 

Moments of the mass operator can also be considered. 
The second moment is 
\begin{align}
    \sum_{j_1,j_2}
    &\,
    \sum_{k_1,k_2}
    X_{j_1,k_1}^4
    \,
    X_{j_2,k_2}^4
\nonumber\\
    \rightarrow&\,
    \sum_{j_1,j_2}
    \sum_{k_1,k_2}
    T_{(j_1 j_1 j_1 j_1)(j_2 j_2 j_2 j_2)}
    \,
    \bar{T}_{(k_1 k_1 k_1 k_1)(k_2 k_2 k_2 k_2)},   
\end{align}
where the second expression is relevant for the non-abelian
bosonization. Here the holomorphic part 
$T_{(j_1 j_1 j_1 j_1)(j_2 j_2 j_2 j_2)}$
is fully symmetrized in two groups of 4 indices.
The most relevant component obtains from subsequently
antisymmetrizing these in pairs.
The corresponding representation has highest weight
$\Lambda = 4 \, \omega_2$. 
This generalizes to the antisymmetrized $p$th moment. 
We note that all such moment operators are primary fields
for generic $R$ and level $q = 8$ \cite{BYB}. 
The scaling dimension of the $p$th moment is 
\begin{align}
    \Delta_{M,p}
    =
    \frac{4 p(R + 3 - p)}{D_q(R)},
\end{align}
where $D_q(R)$ is defined by Eq.~(\ref{WZNWDen}).
This result indicates that moments of the mass operator
are (weakly) \emph{multifractal} at leading (zeroth) order in $\epsilon$. 


\begin{figure}[b!]
\centering
\includegraphics[width=0.25\textwidth]{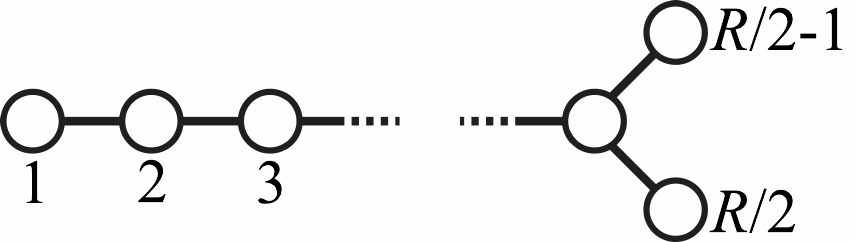}
\caption{Dynkin diagram for the Lie algebra so($R$) with $R$ even (also known as the algebra $D_{R/2}$ in the Cartan classification) \cite{BYB}.  
Nodes are associated to fundamental weight representations. The first node labeled ``$1$'' corresponds to the $R$-dimensional vector representation $V_j$ (weight $\Lambda = \omega_1$). 
The nodes ``$R/2 -1$'' and ``$R/2$''  label the left- and right- spinor representations (weights $\Lambda = \omega_{R/2 -1}$ and $\omega_{R/2}$).
}
\label{Fig--Dn}
\end{figure}

To encode the ``magnetic'' vertex operators $V_{0,m}$ in Eq.~(\ref{SM--VertOp}),
we consider the spinor representations of SO($R$). 
In order to represent vortex operators for the $R = 2$ theory, 
we take $R$ even and examine tensors built from the independent
left- and right-handed spinor representations. These are associated to the fundamental weights $\omega_{R/2 -1}$ and $\omega_{R/2}$, indicated by the labeled nodes in the Dynkin diagram shown in Fig.~\ref{Fig--Dn}.

We can associate fields $\psi_{\sigma}$ and $\zeta_{\sigma'}$ to
the fundamental spinor representations 
$\Lambda = \omega_{R/2 -1}$
and
$\Lambda = \omega_{R/2}$,
respectively. 
The spinor indices $\sigma,\sigma' \in \{1,2,3,\ldots,2^{R/2 - 1}\}$, so that each is $2^{R/2-1}$-dimensional. 
For $R = 4$ [SO(4)], these correspond to 2-component left- and right-handed Weyl fermions.
In the $R \rightarrow 2$ limit,
both $\psi_{\sigma}$ and $\zeta_{\sigma'}$ become \emph{scalars}.
Then we can link a tensor field with $r$- and $s'$-fold symmetrized indices with a holomorphic vertex operator,
\begin{align}\label{rsfield}
    T_{(\sigma_1 \cdots \sigma_{r})(\sigma_1' \cdots \sigma_s')}
    \sim&\,
    \left(T_{\sigma_1}\right)^r
    \left(T_{\sigma_1'}\right)^{s'}
\nonumber\\
    \rightarrow&\,
   : e^{\ii (r - s') \varphi(z) /2} :.
\end{align}
Since $r, s' \geq 0$, we see that both left- and right- representations are needed to get both positive and negative charges $(r - s')/2$.

In the WZNW theory, the field in Eq.~(\ref{rsfield}) carries the holomorphic dimension 
\begin{align}
    h_{r + s'}
    =
    \frac{C_2(r \, \omega_{R/2 - 1} + s' \, \omega_{R/2})}{2 D_q(R)}
    \rightarrow
    \frac{(r - s')^2}{64}.
\end{align}
Here $C_2(\Lambda)$ denotes the quadratic Casimir for the highest-weight representation $\Lambda$, and we take $R \rightarrow 2$ at the end.
At level $q = 8$, we should restrict $r + s' \leq 8$ \cite{BYB}. 
Then the vortex operators $V_{0,\pm 1}$ can be identified with 
\begin{align}
\begin{array}{lll}
    V_{0,1}
    =
    : e^{\ii 4 \theta} :
    &\leftrightarrow&
    T_{\sigma_1 \cdots \sigma_8}(z)
    \,
    \bar{T}_{\sigma_1' \cdots \sigma_8'}(\bar{z}),
\\
    V_{0,-1}
    =
    : e^{- \ii 4 \theta} :
    &\leftrightarrow&
    T_{\sigma_1' \cdots \sigma_8'}(z)
    \,
    \bar{T}_{\sigma_1 \cdots \sigma_8}(\bar{z}).
\end{array}
\end{align} 
Rank-$2 p$ tensors of this type carry the scaling dimension 
\begin{align}\label{SM--spinorDim}
    \Delta_{p}
    =
    \frac{R p \left[p + (R-2)\right]}{8 D_q(R)}.
\end{align}
For $R = 2$ and $p = 8$, we get $\Delta_8 = 2 h_{0,1} = 2$, 
which is the vortex scaling dimension $\Delta_V$ at criticality ($K = K_c = 2/\pi$).

\end{document}